\documentclass[aps,prb,showpacs,twocolumn]{revtex4}
\usepackage{graphicx,dcolumn,bm,amsmath,amssymb}

\begin{document}

\newcommand{\be}{\begin{equation}}
\newcommand{\ee}{\end{equation}}
\newcommand{\bea}{\begin{eqnarray}}
\newcommand{\eea}{\end{eqnarray}}
\newcommand{\br}{{\bf r}}

\title{Interaction-induced dephasing of Aharonov-Bohm oscillations}

\author{T.~Ludwig$^{1}$}
\author{A.~D.~Mirlin$^{1,2,*}$}
\affiliation{$^1$Institut f\"ur Nanotechnologie, Forschungszentrum
 Karlsruhe, 76021 Karlsruhe, Germany\\
\mbox{$^2$Institut f\"ur Theorie der Kondensierten Materie,
Universit\"at Karlsruhe, 76128 Karlsruhe, Germany}}

\date{May 25, 2003}

\begin{abstract}
We study the effect of the electron-electron interaction on the
amplitude of mesoscopic Aharonov-Bohm oscillations in
quasi-one-dimensional (Q1D) diffusive rings. We show that the
dephasing length $L_\phi^{\rm AB}$ governing the damping factor 
$\exp(-2\pi R /L_\phi^{\rm AB})$ of the oscillations
is parametrically different from the common
dephasing length for the Q1D geometry. This is due to the fact that
the dephasing is governed by energy transfers determined by the ring
circumference $2\pi R$, making $L_\phi^{\rm AB}$ $R$-dependent. 
\end{abstract}


\maketitle

The Aharonov-Bohm (AB) oscillations of conductance are one of the most
remarkable manifestations of electron phase coherence in mesoscopic
samples. Quantum interference of contributions of different electron
paths in a ring threaded by a magnetic flux $\Phi$ makes the
conductance $g$ an oscillatory function of $\Phi$, with a period
of the flux quantum $\Phi_0=hc/e$; see
Refs.~\onlinecite{Washburn,Aronov_Sharvin,Imry} for reviews.
In a diffusive ring these $\Phi_0$-periodic conductance oscillations
are sample-specific (and would vanish upon the ensemble averaging),
due to a random phase associated with diffusive paths. In this respect,
the $\Phi_0$-periodic AB effect is a close relative of mesoscopic
conductance fluctuations. 

Another type of the AB effect is induced by interference of
time-reversed paths encircling the ring and is intimately related to
the weak localization (WL) correction \cite{Aronov_Sharvin}. Its
principal period is $\Phi_0/2\,$. It survives the ensemble averaging
but is suppressed by a magnetic field penetrating the sample. Below we
concentrate on the first (mesoscopic, or $\Phi_0$-periodic) AB
effect. Our results are, however, applicable to the second
(weak-localization, or $\Phi_0/2$-periodic) type of AB oscillations as
well, as we discuss at the end of the paper.  

In the present paper we will investigate the effect of the
electron-electron interaction on the amplitude of the AB
oscillations. 
Interaction-induced inelastic processes lead to dephasing of
electrons, and thus to a damping of interference phenomena, in
particular, of AB oscillations. The mesoscopic AB
oscillations can thus serve as a ``measuring device'' for the electron
decoherence. This idea was, in particular, implemented in recent
experiments, \cite{Pierre_Birge_PRL, Birge_2003} where the
low-temperature behavior of the dephasing time $\tau_\phi$ was
studied, and two mechanisms of decoherence were identified: scattering
off magnetic impurities\cite{Falko} and electron-electron scattering.

Let $g(\Phi)$ be the dimensionless (measured in units of $e^2/h$)
conductance of the ring. 
It is convenient to expand the conductance fluctuations $\delta
g(\Phi)$ as a Fourier series
\be
\label{e1}
\delta g(\Phi)=\delta g_0 + 2 \sum _{n=1}^\infty \delta g_n
\cos\left(2\pi n\Phi/\Phi_0+\theta_n\right)\:. 
\ee
Within the conventional approach, when the dephasing time
$1/\tau_{\phi}$ is introduced as  a mass of the diffuson and cooperon
propagators, ${\cal P}_{D,C}(q,\omega)\sim 1/(Dq^2-{\rm i}\omega+1/\tau_\phi)$,
the variance of the $n$th harmonic of the oscillations is suppressed
by the factor \cite{Aronov_Sharvin}
\be
\label{e2}
\left\langle \delta g_n^2\right\rangle \sim \frac{L_T^2\,L_\phi}{R^3} 
e^{-2\pi R n/L_\phi}\:,
\ee
where $L_\phi=(D\tau_\phi)^{1/2}$ is the dephasing length,
$L_T=(D/T)^{1/2}$ the thermal length, $D$ the diffusion constant, 
$T$ the temperature, $R$ the radius of the ring, and we set $\hbar=1$.
[It is assumed in Eq.~(\ref{e2}) that $L_\phi,L_T\ll 2\pi R$.]
For a thin ring, $L_\phi$ is then
identified with the dephasing length governing the WL 
correction in the quasi-one-dimensional (Q1D) geometry, which was
found by Altshuler, Aronov, and Khmelnitskii
\cite{AAK,Altshuler_Aronov} to be 
\be
\label{e3}
L_\phi=\left(D\tau_\phi\right)^{1/2},\qquad \tau_{\phi}^{-1}\sim 
\left(\frac{T}{\nu D^{1/2}}\right)^{2/3}\:.
\ee
In fact, Aleiner and Blanter showed recently \cite{Aleiner_Blanter}
that $\tau_\phi$ relevant to the mesoscopic conductance fluctuations
in wires has indeed the same form, Eq.~(\ref{e3}), as the WL dephasing
time.
This seems to support the assumption that the dephasing times
governing different mesoscopic phenomena are identical. Equations
(\ref{e2}) and (\ref{e3}) are commonly used for the
analysis of experimental data. 

We will show below, however, that contrary to the naive expectations
the formulas (\ref{e2}) and
(\ref{e3}) do not describe correctly the dephasing of AB
oscillations. Specifically, if the interaction-induced exponential
damping factor of AB oscillations is presented in the form 
$\langle\delta g_n^2\rangle\sim\exp(-2\pi R n/L_\phi^{\rm AB})$,
the corresponding length $L_\phi^{\rm AB}$ is parametrically
different from Eq.~(\ref{e3}). Moreover, $L_\phi^{\rm AB}$ depends on
the system size $R$. 

We consider a ring coupled symmetrically by two leads to the bulk
electrodes. We will assume the ring and the leads to be of a Q1D
geometry, i.e.~the widths of the wires are much larger than the
Fermi wavelength but much smaller than $R$ and $L_\phi\,$.
The only geometric parameter characterizing the
problem is then the ratio $\gamma$ of the resistance of the ring
itself to the total resistance of the ring with the leads.
By definition, $0<\gamma<1$. 

Following Refs.~\onlinecite{AAK,Altshuler_Aronov,Aleiner_Blanter,AAG},
the elec\-tron-\-elec\-tron interaction can be represented by external
time-dependent random fields $\varphi^\alpha({\bf r},t)\,$, 
with the correlation function 
$\langle\varphi^\alpha(\br,t)\,\varphi^\beta(\br',t')\rangle$
determined from the fluctuation-dissipation theorem,
\be
\label{e4}
\left\langle\varphi^\alpha({\bf r})\,\varphi^\beta({\bf
r^\prime})\right\rangle_\omega = -{\rm Im}\,U({\bf r},{\bf r^\prime;\omega})\,
\delta_{\alpha\beta}\,{\rm coth}\frac{\omega}{2T}\:.
\ee
The conventional form for the dynamically screened Coulomb interaction
in a diffusive system is \cite{Altshuler_Aronov} 
\be
\label{e5}
U(q,\omega)=\frac{1}{U_0^{-1}(q)+\Pi(q,\omega)}\simeq \Pi^{-1}(q,\omega)\:,
\ee
where $U_0(q)$ is the bare Coulomb interaction, 
$\Pi(q,\omega)= \nu Dq^2/(Dq^2-i\omega)$ is the polarization operator,
and $\nu$ is the density of states.
For a diffusive system, the bare Coulomb interaction may be neglected
compared to the polarization since the inverse screening length,
which is of the order of the Fermi wave number, is much larger than
typical momenta $q$.
As we will see below, the characteristic momenta $q$ for our problem
are of the order of the inverse system size $R^{-1}$. In view of the
nontrivial geometry of our system, it is thus more appropriate to
work in the coordinate representation. A corresponding generalization
of Eq.~(\ref{e5}) can be readily obtained, yielding
\be
\label{e6}
{\rm Im}\,U(\br,\br';\omega)\simeq {\rm Im}\,
\Pi^{-1}(\br,\br';\omega)=-\frac{\omega}{\nu D}{\cal D}(\br,\br')\:,
\ee
where ${\cal D}$ is the propagator for the Laplace equation, 
$-\nabla^2{\cal D}({\bf r},{\bf r^\prime})=\delta({\bf r}-{\bf
r^\prime})\,$ with zero boundary conditions at the contacts with bulk
electrodes.  Substituting Eq.~(\ref{e6}) in Eq.~(\ref{e4}), we get,
for relevant frequencies $\omega\ll T$,
\be
\label{e7}
\left\langle\varphi^\alpha({\bf r},t)\,\varphi^\beta({\bf
r^\prime},t^\prime)\right\rangle = 
\frac{2T}{\nu D}\,{\cal
  D}({\bf r},{\bf r^\prime})\,\delta_{\alpha\beta}\,\delta(t-t^\prime)\:.
\ee

Since the ring we are considering consists of Q1D wires, the
corresponding diffusion propagator satisfies the one-dimensional
diffusion equation
\bea
\label{e8}
\Big\{\partial_t-D\partial_x^2+{\rm
  i}\left[\varphi^\alpha(x,t)-\varphi^\beta(x,t)\right]\Big\}
\,{\cal P}_{\delta\Phi}^{\alpha\beta}(x,t;x',t')\nonumber\\ 
= \delta(x-x')\,\delta(t-t')
\eea
supplemented by appropriate matching conditions at junctions of the
ring and leads. Here $\delta\Phi=\Phi_1-\Phi_2$ is difference
in the AB flux between the two measurements, which is incorporated in
the matching conditions. The conductance correlation function is given
by the conventional two-diffuson diagrams,
\cite{Altshuler_Shklovskii,Kane_Serota_Lee,note1}
yielding (we drop the prefactor of order unity)
\bea
\label{e9}
&& \big\langle\delta g(\Phi_1)\,\delta g(\Phi_2)\big\rangle \sim 
\frac{D^2}{TR^4} \int dx_1\,dx_2 \int dt\,
dt^\prime \nonumber\\ && \qquad
\times\,\tilde{\delta}(t-t^\prime)\,\big\langle
{\cal P}^{12}_{\delta\Phi}(x_1,x_2,t)\,
{\cal P}^{21}_{\delta\Phi}(x_2,x_1,t^\prime)
\big\rangle\:,
\eea
where angular brackets denote averaging over the external fields,
$\tilde{\delta}(t)$ is given by
\bea
\label{e10}
\tilde{\delta}(t)&=&12\pi T\int\frac{d\epsilon_1}{2\pi}
\frac{d\epsilon_2}{2\pi}\,f'(\epsilon_1)\,f'(\epsilon_2)\,
{\rm e}^{{\rm i}(\epsilon_1-\epsilon_2)t}\nonumber\\
&=&3\,\pi\,T^3\,t^2 \,{\rm sinh}^{-2}\left(\pi Tt\right)\:,
\eea
and $f(\epsilon)$ is the Fermi distribution function. The function
$\tilde{\delta}(t)$ is peaked at $t=0$ with a width $T^{-1}$. We will
replace it below by the delta function $\delta(t)$. This is justified
if the dephasing effect during the time $t-t'\sim T^{-1}$ is
negligible, i.e. $\langle \phi^\alpha(x,t)\phi^\alpha(x,t)\rangle
T^{-1} \ll 1$. Using Eq.~(\ref{e7}), we find that the latter condition
is equivalent to the requirement that the conductance of the sample is
much larger than the
conductance quantum $e^2/h \simeq (25\,{\rm k}\Omega)^{-1}$.
This condition is well satisfied in typical experiments with metallic
rings, thus justifying the replacement
$\tilde{\delta}(t)\to\delta(t)$. 

\begin{figure}[h]
\begin{center}
\includegraphics[width=0.92\linewidth]{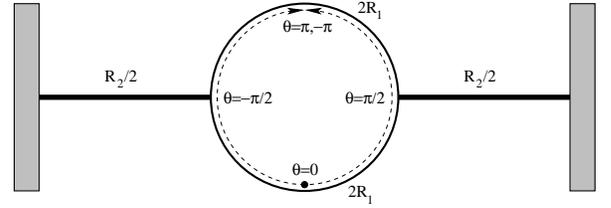}
\caption{\label{ring} The sample geometry and the angular coordinate
  $\theta$ introduced above. The paths representing the saddle-point
  solution are shown. The geometric parameter $\gamma$ is defined as
  the ratio of the resistance of the ring without the leads to the
  resistance of the total sample,
  $\gamma=R_1/(R_1+R_2)\,$.}
\end{center}
\end{figure}

We now express the diffusion propagators in Eq.~(\ref{e9}) as path
integrals. We are interested in the regime of strong dephasing, when
the relevant paths propagate only inside the ring and do not extend
into the leads (see below). It is convenient to introduce the angular
coordinate $\theta$ on the ring ($-\pi\le\theta\le\pi$), with
$\theta=\pm\pi/2$ corresponding to the junctions with the leads
(Fig.~\ref{ring}).
Expanding the
conductance fluctuations in Fourier harmonics with respect to the
flux, $\delta g(\Phi)\to\delta g_n$, we then get
\bea
\label{e11}
\lefteqn{\left\langle\delta g_n^2\right\rangle \sim \frac{D^2}{TR^4} 
\int d\Theta_1\,d\Theta_2 \int dt\,
\int_{\Theta_2}^{\Theta_1}D\theta_1(t)
\int_{\Theta_2}^{\Theta_1}D\theta_2(t)}& &\nonumber\\
&\times&\!{\rm exp}\left\{-\int_0^t
dt^\prime\left[\frac{R^2\dot{\theta_1}^2}{4D}+\frac{R^2\dot{\theta_2}^2}{4D}+V(\theta_1,\theta_2) \right]\right\}\:,
\eea
where the path integral  
goes over pairs of paths $\theta_1(t), \theta_2(t)$ which have a
relative winding number $n$. 
The ``potential'' $V(\theta_1,\theta_2)$ in Eq.~(\ref{e11}) is given by 
$V(\theta_1,\theta_2)=\langle[\phi^\alpha(\theta_1)
-\phi^\alpha(\theta_2)]^2\rangle$; its explicit form can be
straightforwardly obtained according to Eq.~(\ref{e7}) by solving the
diffusion equation in the ring with leads \cite{dipl}.
We will only need below the form of $V(\theta_1,\theta_2)$ for both
coordinates being in the same arm of the ring. For
$|\theta_i| \le \pi/2$ we find
\be
\label{e12}
V(\theta_1,\theta_2)=\frac{2TR}{\nu D}
\left[\left|\theta_1-\theta_2\right|-\frac{\gamma+1}{2\pi}
\left(\theta_1-\theta_2\right)^2\right];
\ee
the expression for $|\theta_i|>\pi/2\,$ follows from symmetry
considerations.

\begin{figure}[h]
\begin{center}
\includegraphics[width=0.92\linewidth]{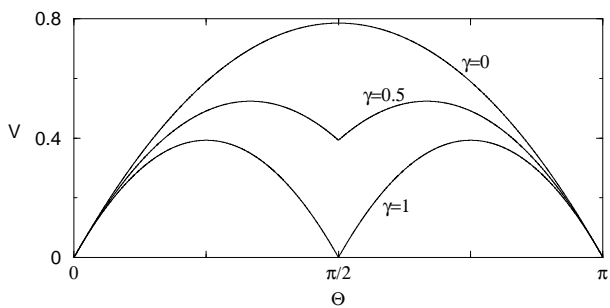}
\caption{\label{potential} The potential $V(\theta)$ of the tunneling
problem, Eq.~(\ref{e13}), plotted in units of $4TR/(\nu D)$ for
different values of $\gamma$.}
\end{center}
\end{figure}

We consider first the fundamental harmonic ($n=1$) of the AB
oscillations; a generalization to higher harmonics, $n=2,3,\ldots$
will be done in the end. For $n=1$ the relevant pairs of paths
interfere after half encircling the ring in the opposite directions. 
We are interested in the regime of a relatively high temperature, when
the dephasing effect is strong.
In this case, the path integral in Eq.~(\ref{e11})
can be evaluated via the saddle-point method. As has been mentioned
above, the paths representing the saddle-point solution (instanton) do
not extend into the leads. Indeed, exploring a part of a lead and
returning back into the ring would only increase the action
of the path.
It is clear from the symmetry considerations that the optimal paths
satisfy $\theta_1(t)=-\theta_2(t)$. Furthermore, it is easy to see
that the optimal initial and final points are located at maximum
distance from the leads, i.e.~$\Theta_1=0$ and $\Theta_2=\pi$ (or vice
versa).
To within exponential accuracy, the problem is then reduced to that
of a particle of mass $R^2/D$ tunneling with zero energy in the
potential (Fig.~\ref{potential})
\be
\label{e13}
V(\theta)=\frac{4TR}{\nu D}\times\left\{
\begin{array}{ll}
\left[\theta-\frac{\gamma+1}{\pi}\theta^2\right]\,,\ \ 
& 0\le\theta\le\frac{\pi}{2} \\
\left[(\pi-\theta)-\frac{\gamma+1}{\pi}(\pi-\theta)^2\right]\,,\ \ 
&\frac{\pi}{2}\le\theta\le\pi
\end{array} \right.
\ee
from $\theta=0$ to $\theta=\pi$. Since the potential is composed of
quadratic parts, the corresponding instanton action is easily
calculated, yielding $\langle \delta g_1^2\rangle \propto e^{-S}$ with
\be
\label{e14}
S = C_\gamma\,\frac{T^{1/2}\,R^{3/2}}{\nu^{1/2}\,D}\:.
\ee
Here $C_\gamma$ is a coefficient of order unity depending on
the geometrical factor
$\gamma$,
\be
\label{e15}
C_\gamma= \left[\frac{\pi}{2(\gamma+1)}\right]^{3/2}
\left[2\gamma\left(1-\gamma^2\right)^{1/2}
+\pi+2\arcsin\gamma\right]\:;
\ee
$C_\gamma$ is equal to $\pi^{5/2}/2^{3/2}$ in the limit of long leads 
($\gamma\to 0$) and to $\pi^{5/2}/4$ in the limit of short leads 
($\gamma\to 1$).
The above calculation can be straightforwardly
generalized to higher harmonics of the AB oscillations,
\mbox{$n=2,3,\ldots\,$}. The optimal paths still begin at $\Theta=0$
or $\pi$ but now perform $n/2$ windings in the opposite directions.
Therefore, the corresponding action is $S_n=nS$.

To calculate the preexponential factor, we have to take into account
small fluctuations of the initial and final points $\Theta_1,
\Theta_2$ around their optimal
values, as well as fluctuations of the paths $\theta_1(t),\theta_2(t)$
around the instanton solution. The corresponding calculation is
sketched in the Appendix. The final result for the variance of the
harmonics  of the mesoscopic  AB oscillations  reads (up to a
numerical prefactor)
\be
\label{e16}
\left\langle\delta g_n^2\right\rangle \sim
\left(\frac{L_T}{R}\right)^{7/2}
\left(\frac{\nu D}{R}\right)^{3/4}{\rm e}^{-nS}\:,
\end{equation}
where $n=1,2,\ldots\:$, 
and the action $S$ is given by Eq.~(\ref{e14}). 

Let us discuss the obtained result (\ref{e16}), (\ref{e14}). 
First of all, it is
essentially different from what one would obtain by using the formulas
(\ref{e2}), (\ref{e3}). Indeed, the exponent in Eq.~(\ref{e16}) scales in a
different way with the temperature and with the system size, as
compared to Eqs.~(\ref{e2}) and (\ref{e3}). It is instructive to write
Eq.~(\ref{e16}) in a form analogous to Eq.~(\ref{e2}),
\be
\label{e17}
\left\langle\delta g_n^2\right\rangle \:\sim\: \left(\frac{L_T}{R}\right)^2
\left(\frac{L_\phi^{\rm AB}}{R}\right)^{3/2}
e^{-2\pi n R/L_\phi^{\rm AB}}\:, 
\ee
thus defining the Aharonov-Bohm dephasing length 
$L_\phi^{\rm AB}$,
\be
\label{e18}
L_\phi^{\rm AB}=\frac{2\pi}{C_\gamma}\,\frac{\nu^{1/2}\,D}{T^{1/2}\,R^{1/2}}\:,
\ee
which is much shorter than $L_\phi$ in the strong-dephasing regime
$L_\phi^{\rm AB}\ll 2\pi R\,$. The corresponding dephasing rate
$1/\tau_\phi^{\rm AB}=D/(L_\phi^{\rm AB})^2$  is thus given by
\be
\label{e19}
\frac{1}{\tau_\phi^{\rm AB}} = \left(\frac{C_\gamma}{2\pi}\right)^2\,
\frac{TR}{\nu D}\:.
\ee
To shed more light on the physical reason for the difference between
the conventional Q1D formula (\ref{e3}) and our result (\ref{e18}),
(\ref{e19}), the following qualitative argument is instructive. 
Calculating perturbatively the dephasing rate using the formula
(\ref{e5}) for the screened Coulomb interaction in a diffusive system,
one gets
\be
\label{e20}
\tau_\phi^{-1}=\int \frac{dq}{2\pi}\,\frac{T}{\nu Dq^2}\:.
\ee
In the calculation of the dephasing rate in a wire
\cite{Altshuler_Aronov,AAG,Aleiner_Blanter}, the arising infrared
divergence is cut off self-consistently, since only processes with
energy transfers $\omega\gtrsim\tau_\phi^{-1}$ contribute. 
As a result, the
lower limit of integration in Eq.~(\ref{e20}) is $\sim L_\phi^{-1}$,
yielding the result (\ref{e3}). On the other hand, in the case of the
Aharonov-Bohm dephasing rate, the relevant paths have to encircle the
ring. Therefore, despite the fact that $L_\phi^{\rm AB}\ll 2\pi R$,
the low-momentum cutoff in Eq.~(\ref{e20}) is set by the inverse
system size $(2\pi R)^{-1}$. This yields
$1/\tau_\phi^{\rm AB}\sim TR/(\nu D)$, reproducing (up to a numerical
coefficient) the result (\ref{e19}).

It is worth emphasizing that our result (\ref{e19}) for the dephasing
rate depends also on the geometry of the leads through the coefficient
$C_\gamma\,$. We note a certain similarity to the dependence of the
dephasing rate in a {\it ballistic} AB-ring on the probe configuration
recently found in Ref.~\onlinecite{Buttiker}.

As has been mentioned in the introduction, our results are also
applicable to the WL ($h/2e$-periodic) AB oscillations. Their $n$th
harmonic $\delta g_n^{\rm WL}$ is determined by cooperon paths with
winding number $n$. Assuming that the magnetic flux penetrating the
sample is negligible and comparing the path-integral representations
for $\langle\delta g_n^2\rangle$ and $\delta g_n^{\rm WL}\,$ we find
\be
\label{e21}
\left\langle\delta g_n^2\right\rangle = \frac{e^2D}{3TL^2}\,\delta
g_n^{\rm WL}\:,
\ee
where $L=\pi R/\gamma$. This implies, in particular, that the
dephasing length $L_\phi^{\rm AB}\,$, Eq.~(\ref{e18}), is the same for
both types of the AB effect. Equation (\ref{e21}) is a generalization
of the relation\cite{Aleiner_Blanter} between the WL correction and
conductance fluctuations for single-connected geometries.

To summarize, we have studied how the Aharonov-Bohm
oscillations are suppressed by dephasing caused by the
electron-electron interaction. Using the path-integral formalism and
the instanton method, we have obtained the result (\ref{e16}),
(\ref{e14}) which is parametrically different from the naive
expectation (\ref{e2}), (\ref{e3}). This demonstrates that the
AB dephasing rate $1/\tau_\phi^{\rm AB}$,
Eq.~(\ref{e19}),  is parametrically different from the dephasing rate
$1/\tau_\phi\,$, Eq.~(\ref{e3}), corresponding to a single-connected
geometry. Physically, the difference can be traced back to the fact
that $1/\tau_\phi$ is determined self-consistently by the processes
with energy transfers of the order of $1/\tau_\phi$ itself
(or equivalently with momentum transfers $\sim 1/L_\phi$),
while the characteristic energy and momentum transfers governing
$1/\tau_\phi^{\rm AB}$ 
are determined by the system size. For this reason, the Aharonov-Bohm
dephasing rate $1/\tau_\phi^{\rm AB}$ depends on the ring radius $R$,
diverging in the limit $R\to\infty\,$.

Valuable discussions with I.L.~Aleiner, B.L.~Altshuler,
N.O.~Birge, H.~Bouchiat, M.~B\"uttiker, V.I.~Falko, and  I.V.~Gornyi 
are gratefully acknowledged.

\appendix*
\section{}

To find  the preexponential factor in (\ref{e16}) for the
variance of the amplitude of the AB oscillations, we have to perform
several Gaussian integrations over deviations from the instanton
solution in the path integral (\ref{e11}). We will only calculate the
parametric dependence of the pre\-factor, neglecting numerical factors
of order unity.

First, let us consider small offsets of the initial and final
points of the paths from their optimal position. The second-order
variation of the 
action $\delta^2S$ will be a quadratic form of the offsets, $\delta^2
S=u_{ij}\,\delta\Theta_i\,\delta\Theta_j$, where $i=1,2$. Using that 
$\delta^2S \sim 1$ for $\delta\Theta_i\sim 1$, we get 
\be
\left({\rm det}\,u_{ij}\right)^{-1/2} \sim S^{-1}\:.
\label{a1}
\ee
Second, we have to account for small deviations of the paths
from the instanton solution. The corresponding factor
can be identified as the propagator for a harmonic oscillator with the
parameters $m\sim R^2/D$ and
$m\omega^2\sim RT/D\nu$. There are two such factors
(one for each of the paths), yielding together
\be
\left[\left(m\omega\right)^{1/2}\right]^2\sim
\frac{T^{1/2}\,R^{3/2}}{\nu^{1/2}\,D}\:. 
\label{a2}
\ee
Finally, there is a Gaussian integration over the deviations of the
time $t$ spent on the path from its optimal value
$t_{\rm opt}\sim (\nu R/T)^{1/2}$.
The corresponding factor can be estimated as 
\be
\label{a3}
\left(\frac{\partial^2S}{\partial t^2}\right)^{-1/2}_{t=t_{\rm
opt}} \sim \left(\frac{S}{t_{\rm opt}^2}\right)^{-1/2} 
\sim \frac{\nu^{3/4}\,D^{1/2}}{T^{3/4}\,R^{1/4}}\:.
\ee
Combining Eqs.~(\ref{e11}), (\ref{a1}), (\ref{a2}) and (\ref{a3}), 
we obtain the
total preexponential factor as given in Eq.~(\ref{e16}).


\begin{thebibliography}{1}
\bibitem{note1} We discard the cooperon contribution to the
correlation function, which does not influence the variance 
$\langle g_n^2\rangle$ of harmonics of the AB oscillations but only
the statistics of the oscillation phase. If the flux threading
the metal is much less than one flux quantum, the phases $\theta_n$ in
Eq.~(\ref{e1}) will be $0$ or $\pi$, while in the opposite case they are
randomized.
\end{thebibliography}
\end{document}